\documentclass[12pt]{article}
\usepackage{float} 
\usepackage{graphicx, subcaption, float, lipsum}
\usepackage{amsfonts,amssymb,epsfig}
\usepackage[latin1]{inputenc}
\usepackage[english]{babel}
\usepackage{color}
\usepackage{hyperref}
\usepackage{amsmath}
\usepackage{amssymb}
\usepackage{graphicx}
\usepackage{subcaption}
\newtheorem{thm}{Th\'eor\`eme}[section]
\newtheorem{cor}[thm]{Corollaire}
\newtheorem{lem}[thm]{Lemme}
\newtheorem{pro}[thm]{Proposition}
\newtheorem{dfn}[thm]{D\'efinition}
\newtheorem{rmq}[thm]{Remark}
\newtheorem{expl}[thm]{Exemple}

\def\dessous#1\sous#2{\mathrel{\mathop{\kern0pt#2}\limits_{#1}}}

\newcommand{\beq}{\begin{eqnarray}}
\newcommand{\eeq}{\end{eqnarray}}
\newcommand{\bpro}{\begin{pro}}
\newcommand{\epro}{\end{pro}}
\newcommand{\blem}{\begin{lem}}
\newcommand{\elem}{\end{lem}}
\newcommand{\bdfn}{\begin{dfn}}
\newcommand{\edfn}{\end{dfn}}
\newcommand{\bcor}{\begin{cor}}
\newcommand{\ecor}{\end{cor}}
\newcommand{\bthm}{\begin{thm}}
\newcommand{\ethm}{\end{thm}}
\newcommand{\bex}{\begin{expl}}
\newcommand{\eex}{\end{expl}}
\newcommand{\brmq}{\begin{rmq}}
\newcommand{\ermq}{\end{rmq}}
\newcommand{\benum}{\begin{enumerate}}
\newcommand{\eenum}{\end{enumerate}}
\newcommand{\bitem}{\begin{itemize}}
\newcommand{\eitem}{\end{itemize}}
\begin{document}
	\setcounter{section}{0}
	\setcounter{equation}{0}
	\setcounter{figure}{0}
	\setcounter{table}{0}
	\setcounter{footnote}{0}

	
	\vspace*{10mm}

	\begin{center}
		{\bf\Large Thermodynamics for an electron gas in a uniform magnetic field in nonextensive {statistics}} 
		\vspace{10pt}
	\end{center}
	
	\begin{center}
		\vspace{10pt}
		
		{{Bienvenu Gnim Adewi $\!^{\rm a}$}, Isiaka Aremua $\!^{\rm a,b}$, Laure Gouba $\!^{\rm c}$
			}
		\\[3mm]
		
		$^{\rm a}$ \textsl{\small Université de Lomé, Faculté des Sciences, Département de Physique, \\
			Laboratoire de Physique des Matériaux et de Mécanique Appliquée, \\
			02 BP 1515 Lomé, Togo} \\[2mm]
		
		$^{\rm b}$ \textsl{\small University of Abomey-Calavi, International Chair in Mathematical Physics \\
			and Applications (ICMPA), 072 B.P. 050 Cotonou, Benin} \\[2mm]
		
		$^{\rm c}$ \textsl{\small The Abdus Salam International Centre for Theoretical Physics (ICTP), \\
			Strada Costiera 11, I-34141 Trieste, Italy} \\[2mm]
		
		E-mail: bienvenuadewi@gmail.com, claudisak@gmail.com, lauregouba@gmail.com
	\end{center}
	
	\vspace{15pt}

	\begin{abstract}
    This work deals with the physical system governed by a Hamiltonian operator,
	in two-dimensional space, of spinless charged particles subject to a perpendicular magnetic field
	{\bf B}, coupled with a harmonic potential in the context  of nonextensive statistical thermodynamics. The thermodynamics of
	such a quantum gas system is elaborated  in the framework of Tsallis statistics by obtaining the  $q$ versions of the partition function, magnetization, 
	and susceptibility after performing the Hilhorst integral transformation. 
	 The results  are discussed in the $q\rightarrow 1$ limit.
	\end{abstract}

	\section{Introduction}
	The system of charged quantum particles in a constant magnetic field is one of the most extensively studied models in quantum mechanics. Its importance arises from its wide-ranging applications in condensed matter physics and quantum optics, where it provides key insights into phenomena like the quantum Hall effect and Landau quantization. This system remains a fundamental tool for exploring quantum behavior and electromagnetic interactions in both theoretical and experimental research. 
	The Landau problem is based on the motion of a charged particle
	in a uniform magnetic field. The particle can only occupy only discrete (or quantum) energy orbitals, called {\it Landau  levels} \cite{L.L}.  The diamagnetism (quantum phenomenon) is observed when the
	magnetization $\mathcal{M}=-\left(\frac{\partial\Gamma}{\partial B}\right)$, $\Gamma$ being  the thermodynamic potential, undergoes fluctuations at given temperatures. Subjected to the external magnetic field, the material develops a weak induced field. Once the external field is removed, this induced magnetization disappears the diamagnetism leads to the determination
	of the magnetic susceptibility $\chi=\frac{\partial \mathcal{M}}{\partial B}$,  which denotes the capability  of a material to be magnetized under the effect of the   magnetic excitation emitted by  this field.
	In these past decades, a great interest has been devoted to the  traditional Boltzmann-Gibbs-Shannon statistical treatment for particular physical systems in which this approach seems to fail  \cite{Plastino}.In
	\cite{Tsallis 2},  Tsallis postulated a new entropy function, after motivated by the difficulties of the standard thermodynamics with nonextensive systems. Since that time, his methods have been the subject of intensive research works, and have found various applications. 
	Tsallis nonextensive  statistics \cite{Germany}, which is a generalized statistical mechanics and contains the
	standard one (Boltzmann-Gibbs), 
	represents   a useful framework for a recently generalized statistical mechanics and thermodynamics, characterized by a real parameter $q$,  the index of nonextensivity, such that for $q=1$  the standard Boltzmann-Gibbs-Shannon statistics is recovered, has played a major contribution in providing new approaches and solutions  \cite{Braz}.   
	Nonextensive statistical mechanics \cite{Abe and Okamoto} has been  developed into an important framework for modeling the thermodynamics of complex systems \cite{Baldovin and  Cerbino} and the information of complex signals \cite{Gell-Mann}.  
	The calculation of thermodynamical quantities in this generalized scheme is more involved than in the usual statistics, frequently making necessary the recourse to numerical approaches \cite{D. Prato}. The Tsallis nonextensive information measure \cite{Gell-Mann} is given by 
		\begin{equation}
			S_{q}=k_{B}\frac{1-\sum_{n=1}^{w}p_{n}^{q}}{q-1}
	\end{equation}
	where $k_{B}$ stands for the  Boltzmann  constant and $\{p_{n}\}$ is a set of   normalized probabilities. 
	Equivalence of  the four versions of the  Tsallis statistics has been discussed in \cite{Ferri and Martinez}, providing  that they can be unified in the sense of deriving the transformations
	that provide a link between any given pair of versions. 
	In \cite{A.Plastino}, aspects of Tsallis's nonextensive thermodynamics as well as magnetic systems that obey this formalism have been studied. Tsallis nonextensive statistics finds its application in information theory. Thus, the role of Tsallis nonextensive information measure in Jaynes' formalism information theory, based on statistical mechanics, is discussed in some detail. This statistic
	seems to give better answers, in many scenarios and explains the behavior of gravity systems by taking into account nonextensivity \cite{ M.Pickup}.  In \cite{Curilef and Pennini}, from the information{\color{red}-}theoretic quantities like the Wehrl entropy and Fisher's information measure,  the
	thermodynamics of the problem leading to Landau's diamagnetism, namely, a free spinless electron in a
	uniform magnetic field, has been studied. As results,  a new Fisher-uncertainty relation, derived via the Cramer-Rao inequality,
	that involves phase space localization and energy fluctuations, has been deduced.
	These scenarios involve disciplines varied by (\cite{E.Lutz}, etc.) 
	Here, our motivation is to redefine the physical quantities as a generalized partition function, magnetization and susceptibility as well as approximately by incorporating the Hilhorst integral transformation in the context of nonextensive thermodynamics, Moreover, to reanalyze the results derived from standard thermodynamic frameworks, extensive research has been conducted to investigate the effects of changing the real parameter $q$ \cite{D. Prato,A.R. Plastino}. A possible connection of the generalized statistical mechanics with quantum groups and its relevance for the Haldane exclusion statistic has been given by Tsallis
	\cite{E. K. Lenzi} 	and Rajagopal \cite{S. Curilef and C. Tsallis}, respectively. The calculation of thermodynamical quantities in this generalized scheme is more involved than in
	the usual statistics, frequently making necessary the recourse to numerical approaches. By employing an integral representation of the gamma function, a connection can be established between the partition function \( Z_q \) for \( q > 1 \). This connection is formally expressed through the \textit{Hilhorst formula}.
	
	The present study focuses on a system of spinless charged particles subjected to a perpendicular magnetic field \( \mathbf{B} \), in conjunction with a harmonic potential described by a Hamiltonian operator. The analysis specifically addresses the transverse motion of electrons in the \( (x, y) \)-plane. Within the framework of Tsallis nonextensive statistics, the thermodynamics of this quantum gas system are rigorously derived and examined, leading to a characterization of \textit{Landau diamagnetism}.\\ The paper is organized as follows. In Section 2, we start with the study of a spinless charged particles gas on the $(x, y)$-space in a magnetic field {\bf B} with an isotropic harmonic potential. In our approach, we use both step and orbit-center coordinate operators to describe the system's dynamics. In Section 3, the Hilhorst integral is applied to investigate the generalized partition function in the context of Tsallis statistics, by considering the  system in the situation of  a thermodynamical equilibrium.
	 		 Section 4 is devoted to  concluding remarks.
\section{The Hamiltonian of  spinless  electrons gas in a uniform magnetic field with
a harmonic potential
}
Consider a gas of  charged particles without spin, restricted to move in a two-dimensional plane $(x, y)$, under the influence of a magnetic field {\bf B} aligned with the $z$-axis. The quantum states and corresponding energy levels of this system were initially explored by Landau \cite{L.L}. By adding a harmonic potential and neglecting Coulomb interactions, the dynamics of this system can be modeled using the Fock-Darwin Hamiltonian \cite{fock,darwin}. The corresponding Hamiltonian of this system is given by
\begin{eqnarray}
\tilde{\mathcal H} &=&\frac{1}{2M}\left({\bf p}+\frac{e}{c}{\bf A}\right)^{2}+\frac{M\omega_{o}^{2}}{2}{\bf r}^{2},
\end{eqnarray}
		where $M$ is the particle mass, $e$  is the particle charge, {\bf p} is the kinetic momentum and {\bf A} is the vector potential. Let us analyze the problem by focusing on the transverse motion of electrons in  the $(x, y)$-space. The single-particle approach used here is valid for systems where electron-electron interactions can be neglected, as demonstrated in studies of Landau diamagnetism \cite{L.L,Isma}.
           Using  the symmetric gauge
		\begin{eqnarray}\label{eq01}
			{\bf A}&=&\left(-\frac{B}{2}y,\frac{B}{2}x\right),
		\end{eqnarray} 
		the classical Hamiltonian takes the form: 
		\begin{eqnarray}
			\tilde{\mathcal H} &=&\frac{1}{2M}\left[\left(P_{x}-\frac{eB}{2c}y\right)^{2}+\left(P_{y}+\frac{eB}{2c}x\right)^{2}\right]+\frac{M\omega_{o}^{2}}{2}\left(x^{2}+y^{2}\right).
		\end{eqnarray}
		Let us achieve the  canonical quantification of this quantum model by introducing the coordinate and moment operators, which are denoted here $R_{i}, P_{i}$, and satisfying
		\begin{eqnarray}
			\left[R_{i},P_{i}\right]&=&i\hbar\delta_{ij}, \qquad \left[R_{_{i}},R_{j}\right]=0,  \qquad \left[P_{i},P_{j}\right]=0,
		\end{eqnarray}
		with $R_{i}=X,Y$ and $P_{i}=P_{x},P_{y}$ , $i=1,2$.  The indices $i,j=1,2$ in the canonical quantization refer strictly to spatial coordinates ($x,y$) and their conjugate momenta, not to particle labels. This is standard notation in 2D systems \cite{feldman1970}.
		 Let's introduce the  convenient change of variables as below:
		\begin{eqnarray}
			Z&=&X+iY,  \qquad	\bar{Z}=X-iY,  \qquad  P_{Z}=\frac{1}{2}\left(P_{x}-iP_{y}\right),  \qquad
			P_{\bar{Z}}=\frac{1}{2}\left(P_{x}+iP_{y}\right),
		\end{eqnarray}
		verifying
		\begin{eqnarray}
			\left[P_{Z},\bar{Z}\right]=0,  \qquad \left[P_{Z},P_{Z}\right]=0,  \qquad \left[\bar{Z},\bar{Z}\right]=0,  \qquad \left[\bar{Z},P_{Z}\right]=0,
		\end{eqnarray} 
		and allowing to take into account the following set of  operators defined by \cite{Albert Feldman, I.Aremua}:
		\begin{eqnarray}\label{feld001}
			\chi_{+}&=&\frac{2i}{M\Omega}P_{\bar{Z}}+\frac{Z}{2}; \qquad \chi_{-}=\frac{-2i}{M\Omega}P_{z}+\frac{\bar{Z}}{2}; \cr
			\pi_{+}&=&M\Omega\left[\frac{Z}{2}-\frac{2i}{M\Omega}P_{\bar{Z}}\right]; \qquad\pi_{-}=M\Omega\left[\frac{\bar{Z}}{2}+\frac{2i}{M\Omega}P_{Z}\right].
		\end{eqnarray}
	The related commutation relations are provided by
		\begin{eqnarray}\label{feld003}
			\left[\pi_{-},\pi_{+}\right]&=&2M\hbar\Omega,  \qquad \left[\chi_{+},\chi_{-}\right]=\frac{2\hbar}{M\Omega}, \qquad \left[\chi_{\pm},\pi_{\pm}\right]=0.
		\end{eqnarray}
 Using (\ref{feld001}) and (\ref{feld003}) together, the classical Hamiltonian $\tilde{\mathcal H}$ turns into its quantum version as follows: 
		\begin{eqnarray}
			 H=\frac{1}{2}\left[\frac{{\pi_{+}}\pi_{-}}{2M}\left(1+\frac{\omega_{c}}{\Omega}\right)+\frac{M\Omega^{2}}{2}\left(1-\frac{\omega_{c}}{\Omega}\right)\chi_{-}\chi_{+}+\hbar\Omega\right],
		\end{eqnarray}
    which can be splited into two parts 
		\begin{eqnarray}
	 H= H^{\left(1\right)}+H^{\left(2\right)},
		\end{eqnarray}
    with
		\begin{equation}
		 H^{\left(1\right)} =\frac{1}{2}\left[\frac{\pi_{+}\pi_{-}}{2M}\left(1+\frac{\omega_{c}}{\Omega}\right)+\hbar\Omega\right], \qquad H^{\left(2\right)}=\frac{1}{2}\left[\frac{M\Omega^{2}}{2}\left(1-\frac{\omega_{c}}{\Omega}\right)\chi_{-}\chi_{+}\right],
		\end{equation}
		respectively, satisfying
		\begin{eqnarray}
        \left[H^{\left(1\right)},H^{\left(2\right)}\right]=0.
		\end{eqnarray}
The operators $\pi_{+}$ and $\pi_{-}$ are respectively
the energy-raising operator  and lowering operator  which act on the eigenstates $|n,m\rangle$ as follows \cite{Albert Feldman,I.Aremua}:
	\begin{equation}
		\pi_{+}|n,m\rangle =\sqrt{2M\hbar\Omega}\sqrt{n+1}|n+1,m+1\rangle, \quad\pi_{-}|n,m\rangle=\sqrt{2M\hbar\Omega}\sqrt{n}|n-1,m-1\rangle.
	\end{equation}
	The operators $\chi_{+}$ and $\chi_{-}$ corresponding to 
	the angular momentum raising operator and lowering operator, respectively, 
 act on the eigenstates $|n,m\rangle$ as follows \cite{Albert Feldman,I.Aremua}:
	\begin{equation}
		\chi_{+}|n,m\rangle = \sqrt{\frac{2\hbar}{M\Omega}}\sqrt{n-m}|n,m\rangle,  \quad\chi_{-}|n,m\rangle = \sqrt{\frac{2\hbar}{M\Omega}}\sqrt{n-m+1}|n,m-1\rangle.
	\end{equation}
Then, the  eigenvalues corresponding to the Hamiltonian $H$ are given by :
	\begin{eqnarray}\label{feld005}
		E_{n,m}&=&\hbar\Omega\left(n+\frac{1}{2}\right)-\frac{\hbar}{2}\left(\Omega-\omega_{c}\right)m,
	\end{eqnarray}
	where $\Omega=\sqrt{\omega_{c}^{2}+4\omega_{o}^{2}}$ and $\omega_{c}=\frac{eB}{Mc}$ the cyclotron frequency. 
	Recall that, in (\ref{feld005}), 
 $n \in \mathbb N$ associated to the energy,
	and $m \in \mathbb Z$ associated to the $z$-projection of the angular momentum, where the constraint $n \geq m$ holds.	\section{Thermodynamical functions analysis by the Hilhorst integral method in Tsallis statistics}
	This paragraph is devoted to deriving relevant thermodynamical quantities from the partition function in Tsallis statistics by application of the Hilhorst integral method. It is worth mentioning  that integral 
	transform formulae  \cite{Solis and Esguerra, HilhorstFormula1, Fa,  D. Prato} 
	have been extensively employed in some previous works to investigate  Tsallis statistical mechanics in various contexts, including  the classical ideal gas \cite{D. Prato}, black body radiation \cite{Mendes}, and two-dimensional self-gravitating systems \cite{Fa}, etc.  Integral transform formulae are of great importance for expressing the Tsallis partition function of a given system in terms of the nonextensivity parameter $q=1$ or Boltzmann-Gibbs partition function.
	\subsection{Calculation of thermodynamical potentials}
	We examine a system in a state of thermodynamic equilibrium that exchanges energy with its environment. Within the framework of Tsallis thermostatistics, the generalized partition function for the canonical ensemble is expressed as follows \cite{EMF1,Plastino4}
	\begin{eqnarray}\label{eq02}
		Z_{q}&=&Tr\left\{\left[1-(1-q)\beta H\right]^{\frac{1}{(1-q)}}\right\},
	\end{eqnarray}
	where $\beta=\frac{1}{k_{B}T}$ and $k_{B}$ being the Boltzmann constant. 
	The generalized free energy $F_{q}$ of the sample \cite{Lavendo1}, using the corresponding  electron density $n$, is obtained from the generalized partition function $Z_{q}$ as \cite{Isma}
	\begin{eqnarray}
		F_{q}=-\frac{n}{\beta}\frac{Z_{q}^{1-q}-1}{1-q}, 
	\end{eqnarray}
	with the generalized magnetization $M_{q}$  delivered by 
	\begin{eqnarray}
		M_{q}=-\frac{\partial F_{q}}{\partial B},
	\end{eqnarray}
	and the generalized susceptibility as 
	\begin{eqnarray}
		\chi_{q}=-\frac{\partial M_{q}}{\partial B}.
	\end{eqnarray}
The Hilhorst transformation \cite{Moran-LopezJL}, along with its extension for \( q < 1 \), establishes a connection between generalized mechanics statistics and standard statistical mechanics. This relationship allows for the straightforward derivation of generalized thermodynamics \cite{D. Prato}.  

Consequently, statistical mechanics can be interpreted as a Hilhorst integral transformation of the corresponding standard quantities, based on the definition of the Gamma function. Starting from the definition of the Gamma function, 
\begin{eqnarray}\label{eq03}
\Gamma\left(\alpha\right)&=&\int_{0}^{\infty}e^{-t}t^{\alpha-1};
\end{eqnarray}
the partition function,  denoted $Z_{q}$, is performed as \cite{Isma}
\begin{eqnarray}\label{eq04}
		Z_{q}&=&\frac{1}{\Gamma\left(\frac{1}{q-1}\right)}\int_{0}^{\infty}\nu^{\frac{1}{1-q}-1}e^{-\nu}e^{\nu(1-q)\beta H}d\nu,
	\end{eqnarray}
where the following change of variables furnished through  (\ref{eq03})
\begin{eqnarray}\label{eq04}
		t&=&\nu\left[1-\left(1-q\right)\beta  H\right],\cr
		\alpha &=&\frac{1}{q-1}, 
			\end{eqnarray}
is achieved.

From (\ref{eq04}), $Z_{q}$ expresses,  by setting $x=\beta\hbar\Omega$ and considering the cylinder of  height $L_{z}$ and radius $R$,  as
	\begin{eqnarray}\label{eq09}
		Z_{q}\left(\beta\right)&=&\frac{1}{\Gamma\left(\frac{1}{q-1}\right)}2\pi R^{2}\frac{M\omega_{c}}{2\pi\hbar}L_{z}\left(\frac{M}{2\pi\hbar^{2}\beta}\right)^{\frac{1}{2}}\left(\frac{1}{q-1}\right)^{\frac{1}{2}}I_{q},
	\end{eqnarray}
	where 
	\begin{equation}
		I_{q} = \int_{0}^{\infty}\nu^{\left(\frac{1}{1-q}-\frac{1}{2}\right)-1}e^{-\nu}\frac{e^{-\frac{\nu x}{2}\left(q-1\right)}}{1-e^{-\nu x\left(q-1\right)}}\frac{1}{1-e^{-\nu(q-1)\frac{x}{2\Omega}\left(\Omega-\omega_{c}\right)}}d\nu,
	\end{equation}
which can be reset as follows
\begin{eqnarray}\label{eq10}
		I_{q}&=&\Gamma\left(\frac{1}{q-1}-\frac{1}{2}\right)\sum_{n=0}^{\infty}\sum_{m=0}^{\infty}\left\lbrace 1+\left(q-1\right)xn+\left(q-1\right)\frac{x}{2\Omega}\left(\Omega-\omega_{c}\right)m  \right\rbrace^{-\left(\frac{1}{q-1}-\frac{1}{2}\right)}.
	\end{eqnarray}
Performing the following identity \cite{Isma}
\begin{eqnarray}
    \Gamma\left(a,b\right)=\frac{\Gamma\left(a-b\right)}{\Gamma(a)}e^{b\log(a)}, \quad a = \frac{1}{q-1}, \, b = \frac{1}{2},
	\end{eqnarray}
	which leads to 
	\begin{eqnarray}\label{eq11}
    \Gamma\left(\frac{1}{q-1}-\frac{1}{2}\right)&=&\Gamma\left(\frac{1}{q-1},\frac{1}{2}\right)\Gamma\left(\frac{1}{q-1}\right)\left(q-1\right)^{\frac{1}{2}},
	\end{eqnarray}
with  $V=2\pi R^{2}L_{z}$, and 
by substituting Eq.(\ref{eq10}) and Eq.(\ref{eq11}) into (\ref{eq09}), one obtains 
\begin{eqnarray}\label{eq06}
	Z_{q}\left(\beta\right)&=&  V\left(\frac{M\omega_{c}}{2\pi\hbar}\right)\left(\frac{M}{2\pi\hbar^{2}\beta}\right)^{\frac{1}{2}}\Gamma\left(\frac{1}{q-1},\frac{1}{2}\right)\frac{1}{1-e^{-\left(\frac{3-q}{2}\right)x}}\frac{1}{1-e^{-\left(\frac{3-q}{2}\right)\frac{x}{2\Omega}\left(\Omega-\omega_{c}\right)}}.	
\end{eqnarray}
Finally, (\ref{eq06}) is written for different values of the parameter $q$ as follows:
\begin{align}
	Z_{q}\left(\beta\right) &= V\left(\frac{M\omega_{c}}{2\pi\hbar}\right)\left(\frac{M}{2\pi\hbar^{2}\beta}\right)^{\frac{1}{2}}\frac{1}{1-e^{-\left(\frac{3-q}{2}\right)x}}\frac{1}{1-e^{-\left(\frac{3-q}{2}\right)\frac{x}{2\Omega}\left(\Omega-\omega_{c}\right)}} \nonumber \\
	&\quad \times \left\{
	\begin{array}{rcl}
    \frac{\Gamma\left(\frac{1}{q-1}-\frac{1}{2}\right)}{(q-1)^{\frac{1}{2}}\Gamma\left(\frac{1}{q-1}\right)} & \text{ } q>1 \\
		\frac{\Gamma\left(\frac{1}{1-q}+1\right)}{(1-q)^{\frac{1}{2}}\Gamma\left(\frac{1}{1-q}+\frac{3}{2}\right)} & \text{ } q<1 
	\end{array}
	\right. 
.\end{align}
When $q\rightarrow 1$, $a\rightarrow\infty$, $\displaystyle{\lim_{a\rightarrow+\infty}\Gamma\left(a,b\right)=1}$,  then (\ref{eq06}) turns to the standard partition function, that is, 
	\begin{eqnarray}\label{eq15}
		Z_{1}\left(\beta\right)&=&  V\left(\frac{M\omega_{c}}{2\pi\hbar}\right)\left(\frac{M}{2\pi\hbar^{2}\beta}\right)^{\frac{1}{2}}\frac{1}{1-e^{-x}}\frac{1}{1-e^{-\frac{x}{2\Omega}\left(\Omega-\omega_{c}\right)}}.
	\end{eqnarray}
	 From the definition of the internal energy {\cite{S.Martinez and H. Vucetich}}
	\beq
	U_{q}&=&-\frac{\partial \ln Z_{q}}{\partial \beta},
	\eeq
	since 
	\beq
	\ln Z_{q}&=&\ln V+\ln \left(\frac{M\omega_{c}}{2\pi\hbar}\right)+\frac{1}{2}\ln\left(\frac{M}{2\pi\hbar^{2}\beta}\right)+\ln\left[ \frac{\Gamma\left(\frac{1}{q-1}-\frac{1}{2}\right)}{(q-1)^{\frac{1}{2}}\Gamma\left(\frac{1}{q-1}\right)}\right]\cr\cr
	&&+\ln\left[\frac{1}{1-e^{-\left(\frac{3-q}{2}\right)x}}\right]+\ln\left[\frac{1}{1-e^{-\left(\frac{3-q}{2}\right)\frac{x}{2\Omega}\left(\Omega-\omega_{c}\right)}}\right],
	\eeq
	we obtain
	\beq
	U_{q}&=&\frac{1}{2\beta}+\left(\frac{3-q}{2}\right)\hbar\Omega\left(\frac{e^{-\left(\frac{3-q}{2}\right)x}}{1-e^{\left(\frac{3-q}{2}\right)x}}+\frac{\left(\Omega-\omega_{c}\right)}{2\Omega}\frac{e^{-\left(\frac{3-q}{2}\right)\frac{x}{2\Omega}\left(\Omega-\omega_{c}\right)}}{1-e^{-\left(\frac{3-q}{2}\right)\frac{x}{2\Omega}\left(\Omega-\omega_{c}\right)}}\right)
	.\eeq
	When $q\rightarrow 1$, we get 
	\beq
	U_{1}&=&\frac{1}{2\beta}+\hbar\Omega\left(\frac{e^{-x}}{1-e^{-x}}+\frac{\left(\Omega-\omega_{c}\right)}{2\Omega}\frac{e^{-\frac{x}{2\Omega}\left(\Omega-\omega_{c}\right)}}{1-e^{-\frac{x}{2\Omega}\left(\Omega-\omega_{c}\right)}}\right).
	\eeq
	The deformed heat capacity  $C_{q}$ is defined in terms of the internal energy $ U_{q}$ by {\cite{S.Martinez and H. Vucetich}}
	\beq
	C_q &=&\frac{\partial U_q}{\partial T} = -k_{B}\beta^{2} \frac{\partial U_q}{\partial \beta}.
	\eeq
	In our case, we obtain the following expression 
	\beq
C_q = \frac{k_B}{2} & - & k_B \beta^2\left( \frac{3-q}{2} \right)^2(\hbar \Omega)^2\Bigg\{    \frac{ e^{-y}}{(1 - e^{-y})^2}+\frac{(\Omega - \omega_c)^2}{4 \Omega^{2}}\frac{e^{-z}}{(1 - e^{-z})^2} \Bigg\},
	\eeq
where 
\beq
y&=&\left(\frac{3-q}{2}\right)\beta\hbar\Omega,\quad z=\left(\frac{3-q}{2}\right)\beta\hbar\frac{\left(\Omega-\omega_{c}\right)}{2}.
\eeq 
When $q\rightarrow 1$, the corresponding heat capacity becomes 
\beq
C_q &\rightarrow& C_1 = \frac{k_B}{2} - k_B \beta^2(\hbar \Omega)^2 \left\{   \frac{ e^{-\beta \hbar \Omega}}{(1 - e^{-\beta \hbar \Omega})^2} + \frac{ (\Omega - \omega_c)^2}{4 \Omega^{2}} \frac{e^{-\frac{\beta \hbar (\Omega - \omega_c)}{2}}}{(1 - e^{-\frac{\beta \hbar (\Omega - \omega_c)}{2}})^2} \right\}.
\eeq
	Next, we deduce the free energy from its definition\cite{Isma}
	\begin{eqnarray}\label{eq12}
		F_{q}&=&\frac{-n}{\beta}\frac{Z_{q}^{1-q}-1}{1-q}=-nk_{B}T\ln_{q}Z_{q},
	\end{eqnarray}
providing
	\begin{eqnarray}
		F_{q}&=&-\frac{n}{\beta\left(1-q\right)}\left\lbrace \left(\frac{eVB}{2\pi\hbar c}\right)^{1-q}\left(\frac{M}{2\pi\hbar^{2}\beta}\right)^{\frac{1-q}{2}}\Gamma\left(\frac{1}{q-1},\frac{1}{2}\right)^{1-q}\right.\cr
		&&\left.\times\left(\frac{1}{1-e^{-\left(\frac{3-q}{2}\right)\beta\hbar\Omega}}\frac{1}{1-e^{-\left(\frac{3-q}{2}\right)\frac{\beta\hbar}{2}\left(\Omega-\omega_{c}\right)}}\right)^{1-q}-1 \right\rbrace.
	\end{eqnarray}
	Besides, the deformed magnetization is provided  by its definition
	\begin{eqnarray}
		M_{q}&=&-\frac{\partial F_{q}}{\partial B}, 
	\end{eqnarray}
	 as follows 
	 \begin{eqnarray}\label{eq13}
			M_{q} &=& \frac{n}{\beta}\left(\frac{M}{2\pi\hbar^{2}\beta}\right)^{\frac{1-q}{2}} \Gamma\left(\frac{1}{q-1}, \frac{1}{2}\right)^{1-q}\left(\frac{eV}{2\pi\hbar c}\right)^{1-q}  \left\{ \frac{1}{B^{q}} \left(\frac{1}{1 - e^{-\left(\frac{3-q}{2}\right)\frac{\beta\hbar}{2}(\Omega - \omega_{c})}} \frac{1}{1 - e^{-\left(\frac{3-q}{2}\right)\beta\hbar\Omega}}\right)^{1-q} \right.\cr
			&& + B^{1-q} \left[\left(1 - e^{-\left(\frac{3-q}{2}\right)\frac{\beta\hbar}{2}(\Omega - \omega_{c})}\right) \left(1 - e^{-\left(\frac{3-q}{2}\right)\beta\hbar\Omega}\right)\right]^{q}\left(\frac{3-q}{2}\right)\beta\hbar\left[\frac{e^{2}B}{M^{2}c^{2}\Omega} \frac{e^{-\left(\frac{3-q}{2}\right)\beta\hbar\Omega}}{\left(1 - e^{-\left(\frac{3-q}{2}\right)\beta\hbar\Omega}\right)^{2}}\right.\cr 
			&&\times\left.\frac{1}{1 - e^{-\left(\frac{3-q}{2}\right)\frac{\beta\hbar}{2}(\Omega - \omega_{c})}} + \frac{1}{2}\left(\frac{e^{2}B}{M^{2}c^{2}\Omega} - \frac{e}{Mc}\right)\left. \left.  \frac{e^{-\left(\frac{3-q}{2}\right)\frac{\beta\hbar}{2}(\Omega - \omega_{c})}}{\left(1 - e^{-\left(\frac{3-q}{2}\right)\frac{\beta\hbar}{2}(\Omega - \omega_{c})}\right)^{2}}\frac{1}{1 - e^{-\left(\frac{3-q}{2}\right)\beta\hbar\Omega}} \right]  \right.\right\}.
	\end{eqnarray}
	From the magnetization, we derive the magnetic susceptibility given by 
	\begin{eqnarray}
		\chi_{q}=-\frac{\partial M_{q}}{\partial B}, 
	\end{eqnarray}
	and obtain
\begin{eqnarray}\label{eq14}
\chi_{q}&=&-\frac{n}{\beta}\left(\frac{M}{2\pi\hbar^{2}\beta}\right)^{\frac{\left(1-q\right)}{2}}\left(\frac{eV}{2\pi\hbar c}\right)^{\left(1-q\right)}\Gamma\left(\frac{1}{q-1},\frac{1}{2}\right)^{\left(1-q\right)}\cr
&&\times\left\lbrace -qB^{-\left(1+q\right)}\left(\frac{1}{A_{q}\left(\Omega\right)C_{q}\left(\Omega\right)}\right)^{1-q}+\frac{\beta\hbar B^{1-q}}{2M^{2}c^{2}\Omega}(3-q)\frac{\left(A_{q}\left(\Omega\right)C_{q}\left(\Omega\right)\right)^{q}e^{2}(1-A_{q}(\Omega))}{A_{q}\left(\Omega\right)C_{q}(\Omega)}\right.\cr
&&+\left.\frac{e^{2}(1-C_{q}\left(\Omega\right))}{2M^{2}c^{2}\Omega A_{q}^{2}(\Omega)C_{q}(\Omega)}+\frac{e^{2}B^{1-q}\hbar\beta(1-q)(3-q)}{2M^{2}c^{2}\Omega}\frac{\left[A_{q}(\Omega)C_{q}(\Omega)\right]^{q}\left(1-A_{q}(\Omega)\right)}{A_{q}(\Omega)C_{q}^{2}(\Omega)}\right.\cr
&&\left.+\frac{\left(1-C_{q}(\Omega)\right)}{2}\frac{\left(-\frac{e}{Mc}+\frac{Be^{2}}{M^{2}c^{2}\Omega}\right)}{A_{q}^{2}(\Omega)C_{q}(\Omega)}\right\rbrace,		
\end{eqnarray}

	with $ A_{q}\left(\Omega\right)=\left[1-e^{-\left(\frac{3-q}{2}\right)\beta\hbar\Omega}\right]$, $ C_{q}\left(\Omega\right)=\left[1-e^{-\left(\frac{3-q}{4}\right)\beta\hbar\left(\Omega-\omega_{c}\right)}\right]$. 
	When $q\rightarrow 1$, the quantities derived in (\ref{eq12}),(\ref{eq13}) and (\ref{eq14}) write :
	\begin{eqnarray}\label{thermal007}
		F_{1} &=& -\frac{n}{\beta}\ln Z_{1}\left(\beta\right)=-\frac{n}{\beta}\ln  \left\lbrace V\left(\frac{M\omega_{c}}{2\pi\hbar}\right)\left(\frac{M}{2\pi\hbar^{2}\beta}\right)^{\frac{1}{2}}\frac{1}{1-e^{-\beta\hbar\Omega}}\frac{1}{1-e^{-\frac{\beta\hbar}{2}\left(\Omega-\omega_{c}\right)}}\right\rbrace,\cr
		M_{1} &=&\frac{n}{\beta}\left\lbrace \frac{1}{B}+\left(1-e^{-\frac{\beta\hbar}{2}\left(\Omega-\omega_{c}\right)}\right)\left(1-e^{-\beta\hbar\Omega}\right)\beta\hbar\left[\frac{e^{2}B}{M^{2}c^{2}}\frac{e^{-\beta\hbar\Omega}}{\left(1-e^{-\beta\hbar\Omega}\right)^{2}}\frac{1}{1-e^{-\frac{\beta\hbar}{2}\left(\Omega-\omega_{c}\right)}}\right. \right.\cr
		&&\left. \left. +\frac{1}{2}\left(\frac{e^{2}B}{M^{2}c^{2}\Omega}-\frac{e}{Mc}\right)\frac{e^{-\frac{\beta\hbar}{2}\left(\Omega-\omega_{c}\right)}}{\left(1-e^{-\frac{\beta\hbar}{2}\left(\Omega-\omega_{c}\right)}\right)^{2}}\frac{1}{1-e^{-\beta\hbar\Omega}}\right]\right\rbrace,\cr
	   \chi_{1} &=&-\frac{n}{\beta}\left\lbrace -\frac{1}{B^{2}} +\frac{e^{2}(1-A_{1}(\Omega))}{M^{2}c^{2}\Omega}+\frac{e^{2}(1-C_{1}(\Omega))}{2M^{2}c^{2}\Omega A_{1}^{2}(\Omega)C_{1}(\Omega)}+\frac{1-C_{1}(\Omega)}{2}\right.\cr
	&&\left.\times\frac{-\frac{e}{Mc}+\frac{Be^{2}}{M^{2}c^{2}\Omega}}{A_{1}^{2}\left(\Omega\right)C_{1}(\Omega)}\right\rbrace,
\end{eqnarray}
where $ A_{1}\left(\Omega\right)=\left[1-e^{-\beta\hbar\Omega}\right]$, $ C_{1}\left(\Omega\right)=\left[1-e^{-\left(\frac{\beta\hbar}{2}\right)\left(\Omega-\omega_{c}\right)}\right]$.
\begin{figure}[H]
	\centering
	\begin{subfigure}[b]{0.45\textwidth}
		\includegraphics[width=\textwidth]{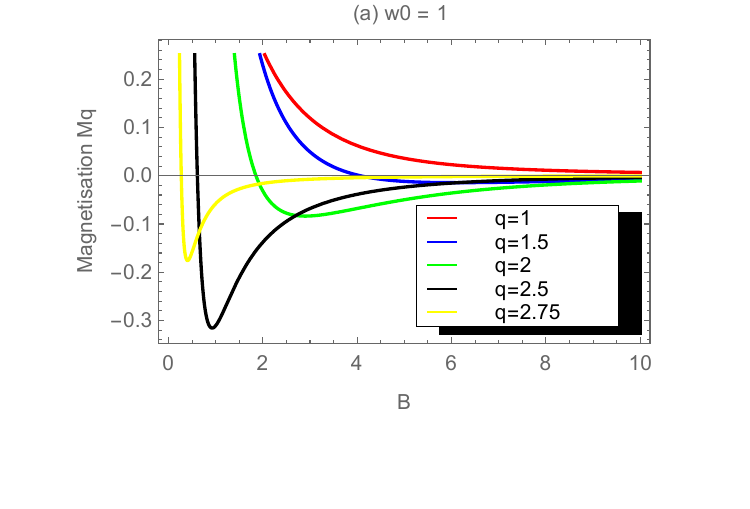}
	\end{subfigure}
	\hfill
	\begin{subfigure}[b]{0.45\textwidth}
		\includegraphics[width=\textwidth]{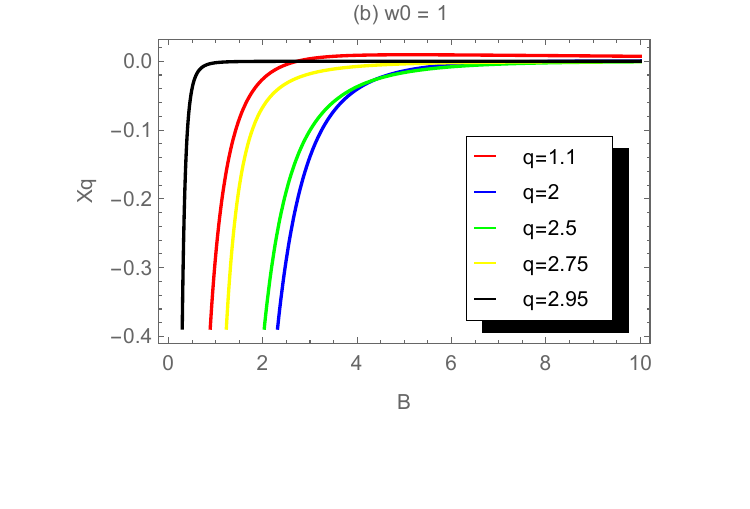}
	\end{subfigure}
	\caption{\it \small  
		Plots of the   deformed magnetization (a) and the magnetic susceptibility (b) of Hilhorst integral method versus  $ B$ :  for  different  values of the deformed parameter $q$.}
\end{figure}
In Figure 1, the graphics display  the behavior of the magnetization $M_q$ and the susceptibility $\chi_q$ versus the intensity  $B$  of the magnetic field  for different values of the nonextensivity parameter $q$:
\subsubsection{Case of the  magnetization}
\begin{itemize}
	\item  For \( q = 1 \):  The magnetization curve is strictly positive and decreases monotonically. These curves  correspond to the standard behavior of magnetization in the Boltzmann-Gibbs statistics framework \cite{R.M.white}. $M_q$ asymptotically approaches zero as \( B \) increases, which is consistent with the expected behavior for a diamagnetic system. For \( q = 1 \), the magnetization tends to zero as \( B \) increases, which is typical of a diamagnetic system. Magnetization can  exhibit more complex behaviors, with transitions between diamagnetic and paramagnetic regimes\cite{C.Kittel,S. Blundell}.
	\item  For \( q \neq 1 \): The magnetization  curves show different behaviors depending on the value of \( q \). For some values of this parameter, $M_q$ indicating a change in magnetic behavior (e.g., a transition from diamagnetism to paramagnetism), the shape of the curves strongly depends on the value of \( q \), highlighting the effect of nonextensivity on the magnetic properties of the system \cite{Tsallis2009}. 
\end{itemize}
\subsubsection{Case of the  susceptibility}
\begin{itemize}
	 \item For \( q = 1 \), the magnetic susceptibility is anticipated to follow a predictable pattern, often characterized by the Curie-Weiss law for paramagnetic materials or a consistent negative value for diamagnetic materials. The negative values observed in the data (-0.1, -0.2, -0.3, -0.4) indicate a diamagnetic response, where the material weakly resists the applied magnetic field \cite{Stanley1971}. If the x-axis represents temperature, the susceptibility may exhibit minor fluctuations while staying negative and small. In diamagnetic materials, susceptibility is typically independent of temperature, so any notable changes could suggest the presence of impurities or secondary phases. A diamagnetic response at \( q = 1 \) implies that the material lacks unpaired electrons or has a fully paired electronic configuration, leading to a weak, negative response to an external magnetic field.
	 \item For values of \( q \neq 1 \) (e.g., \( q = 1.1, 2, 2.5, 2.75, 2.95 \)), the system may display non-trivial behavior, potentially signaling a deviation from conventional magnetic properties. 
	 As \( q \) increases, the susceptibility becomes more negative, suggesting an enhancement of diamagnetic properties \cite{Tsallis2009}. This could be due to changes in the electronic structure, such as increased electron pairing or the formation of localized magnetic moments that oppose the external field more strongly. The curves for different \( q \) values might reveal critical points or phase transitions. For instance, a sharp change in susceptibility at a specific \( q \) value could indicate a transition from a paramagnetic to a diamagnetic state or the onset of a new magnetic phase \cite{Stanley1971}.  The parameter \( q \) is often associated with nonextensive thermodynamics, where \( q \neq 1 \) indicates deviations from standard Boltzmann-Gibbs statistics. In this context, the variation in susceptibility with \( q \) could reflect the influence of long-range interactions, non-equilibrium states, or fractal structures in the material. The increase in diamagnetic response with \( q \) suggests that the material becomes more resistant to magnetization as \( q \) grows. This could be due to the formation of complex magnetic domains or the influence of external factors such as pressure or doping, which alters the electronic properties of the material \cite{A. H. Morrish}. 
	\end{itemize}
	
%
%
	
	
%
\brmq
\begin{enumerate}
	\item For \textbf{\( q \neq 1 \)}, nonextensivity introduces additional effects that modify the magnetic response of the system. These effects can be interpreted as a consequence of long-range interactions or complex correlations between particles, which are not accounted for in standard statistics. The transition between positive values of magnetization suggests  that nonextensivity can induce magnetic phase transitions in the system \cite{A. H. Morrish}. In the limit \( q \to 1 \), the results coincide with those of Boltzmann-Gibbs statistics, where the magnetization is always positive. 
	For \( q = 1 \), the behavior  conforms to the expectations of standard physics.
$M_q$ decreases  with \( B \), and these results demonstrate that Tsallis thermodynamics provides a useful framework for studying complex systems where long-range interactions and correlations play an important role.
	\item \textbf{Comparison between \( q = 1 \) and \( q \neq 1 \)}: the case \( q = 1 \) represents a baseline behavior. The increasing negativity of the susceptibility with \( q \) suggests a strengthening of diamagnetic properties, which could have implications for the material's applications in magnetic shielding or superconductivity \cite{R.M.white, S. Blundell}. The magnetic susceptibility data reveal a clear distinction between the cases \( q = 1 \) and \( q \neq 1 \). While \( q = 1 \) corresponds to a standard diamagnetic response, \( q \neq 1 \) introduces complex behavior that could be linked to nonextensive thermodynamics or changes in the material's electronic structure \cite{Tsallis2009, R.K.Pathria}. 
	\end{enumerate}
\ermq
	\subsection{Discussions at low and high temperatures }
This paragraph discusses the  behavior of the calculated thermal quantities investigated. We first consider the asymptotic expressions at low temperatures, and then at high temperatures. 
		{\subsubsection{Case of low temperature limit} }  Let us consider the situation $x=\beta\hbar\Omega\rightarrow \infty$ implying $\frac{1}{1-e^{-\frac{x}{2\Omega}\left(\Omega-\omega_{c}\right)}}\approx 1+e^{-\frac{x}{2\Omega}\left(\Omega-\omega_{c}\right)}$.
			By substituting these approximations in (\ref{eq15}), the partition function behaves as follows 
			\beq
			Z_{1}\left(\beta\right)\approx V\left(\frac{M\omega_{c}}{2\pi\hbar}\right) \left(\frac{M}{2\pi\hbar^{2}\beta}\right)^{\frac{1}{2}}\left(1+e^{-x}\right)\left(1+e^{-\frac{x}{2\Omega}\left(\Omega-\omega_{c}\right)}\right).
			\eeq
			 Since the  exponentials $e^{-\frac{x}{2\Omega}\left(\Omega-\omega_{c}\right)}$ become negligible at very low temperatures, it comes 
        \beq\label{asymbehav000}
			Z_{1}\left(\beta\right)\approx V\left(\frac{M\omega_{c}}{2\pi\hbar}\right) \left(\frac{M}{2\pi\hbar^{2}\beta}\right)^{\frac{1}{2}}.
			\eeq
			Eq. (\ref{asymbehav000}) shows that $Z_{1}\left(\beta\right)$ grows as $\beta^{-\frac{1}{2}}$ or  $T^{\frac{1}{2}}$
			at low temperatures, only the lowest-energy states are populated, and the partition function is dominated by the lowest-energy term.
			$T^{\frac{1}{2}}$ dependence reflects low-temperature quantum effects. Besides, 
			using the exponential approximations,  the partition  function, $Z_{q}\left(\beta\right)$  
			is dominated by the ground state, so $\ln Z_q$ tends towards a constant. As a result, from (\ref{thermal007}),  for $q\rightarrow 1$, $F_{1}$ tends towards the energy of the ground state and $M_{1}$ turns into
				\beq
			M_{1}\approx \frac{n}{\beta B}.
			\eeq 
			Accordingly, $\chi_{1}$ changes to
			\beq\label{eq16}
			\chi_{1}\approx \frac{n}{\beta B^{2}}.
			\eeq
		
	 {\subsubsection{Case of high temperature limit} 
	 In this context, we have $T\rightarrow\infty,\beta\rightarrow 0$, such that $x=\beta\hbar\Omega\rightarrow 0$, which leads to the  two  following approximations $\frac{1}{1-e^{-x}}\approx \frac{1}{x}=\frac{1}{\beta\hbar\Omega}$ and $\frac{1}{1-e^{-\frac{x}{2\Omega}\left(\Omega-\omega_{c}\right)}}\approx \frac{2\Omega}{x\left(\Omega-\omega_{c}\right)}$.
	Thus, from (\ref{eq15}), we get
	\beq
	Z_{1}\left(\beta\right)\approx V\left(\frac{M\omega_{c}}{2\pi\hbar}\right) \left(\frac{M}{2\pi\hbar^{2}\beta}\right)^{\frac{1}{2}}\frac{2}{\left(\beta\hbar\right)^{2}\Omega\left(\Omega-\omega_{c}\right)}.
	\eeq
It comes  that $Z_{1}\left(\beta\right)$ decreases as $\beta^{-\frac{3}{2}}$ or $T^{\frac{3}{2}}$ at high temperature, the excited states are significantly populated, and the partition function is given by the thermal contributions. The $T^{\frac{3}{2}}$ dependence reflects the classical behavior of an ideal gas. 

Besides for $x=\beta\hbar\Omega$  very small $\left(x\rightarrow 0\right)$,  the partition function $Z_{1}\left(\beta\right)$ typically grows as temperature increases, implying that  $\ln Z_{1}\left(\beta\right)$ becomes large. Thus, the free energy  behaves as 
	\beq
	F_{1}&=&-\frac{n}{\beta}\ln Z_{1}\left(\beta\right)\rightarrow-\infty \left( T \rightarrow +\infty\right),
	\eeq 
highlighting that the system becomes highly disordered at high temperature.

The magnetization $M_{1}$  in this context, becomes 
\beq
M_{1}\approx \frac{n}{\beta}\left\lbrace  \frac{1}{B}+\frac{e^{2}B}{M^{2}c^{2}\Omega}\left(1-x\right)+\frac{\Omega-\omega_{c}}{2\Omega}\left(\frac{e^{2}B}{M^{2}c^{2}\Omega}-\frac{e}{Mc}\right)\frac{2\Omega-x\left(\Omega-\omega_{c}\right)}{\Omega-\omega_{c}}\right\rbrace
.\eeq
 Showing that it can have a complex temperature dependence, but in general, the magnetization decreases at high temperatures due to thermal fluctuations. Thereby, the susceptibility $\chi_{1}$ transforms into 
 \beq
 \chi_{1}&=&-\frac{n}{\beta}\left\lbrace -\frac{1}{B^{2}}+\frac{e^{2}\left(1-x\right)}{M^{2}c^{2}\Omega}+\frac{e^{2}}{M^{2}c^{2}x^{3}\left(\Omega-\omega_{c}\right)}+\frac{1}{2x^{2}}\left(-\frac{e}{Mc}+\frac{Be^{2}}{M^{2}c^{2}\Omega}\right)\right\rbrace.
 \eeq

}
	\section{Concluding remarks}
		In this work, we investigated the Fock-Darwin Hamiltonian describing a gas of spinless charged particles, subject to a perpendicular magnetic field {\bf B}, confined
		in a harmonic potential. We used a set of step and orbit-center coordinate operators which  allowed to obtain the eigenvalues and eigenfunctions of the
		quantum Hamiltonian. Then, we studied the thermodynamics of the physical model by deriving the relevant quantities in the Tsallis nonextensive
		thermodynamics which generalizes and contains the Boltzmann-Gibbs standard one.  From  the $q$-deformed version of the partition function performed
		using the Gamma function in the Hilhorst integral method, by considering a cylindrical body  of volume $V$, radius $R$ and height $L_{z}$, along the $z$-axis,
		containing the gas of charged particles, the corresponding $q$-analog expressions of the internal energy and heat capacity at constant
		volume were determined. In addition, the free energy, the magnetization, and the susceptibility have been achieved. These calculated thermal $q$-quantities 
		were discussed at low and high temperatures, respectively, from the asymptotic behavior of their expressions, which then coincide with the standard ones of
		the Boltzmann-Gibbs statistics in the  $q\rightarrow1$ limit.

\section{Appendix}
This section is devoted to the details of deriving the $q$-deformed partition function  $Z_q$ in (\ref{eq04}) that describes our quantum Hamiltonian. 

Starting from the definition, we have 
\begin{eqnarray}\label{part000}
		Z_{q}&=&\frac{1}{\Gamma\left(\frac{1}{q-1}\right)}\int_{0}^{\infty}\nu^{\frac{1}{1-q}-1}e^{-\nu}e^{\nu(1-q)\beta H}d\nu{\color{blue},}
	\end{eqnarray}
where the denominator of (\ref{part000}), related to the  Hamiltonian $H$,  takes an  explicit expression as follows
\beq
\Gamma\left(\frac{1}{q-1}\right)&=&\int_{0}^{\infty} e^{-\nu\left[1-(1-q)\beta H\right]}  \left[\nu\left[1-(1-q)\beta H\right]\right]^{\frac{1}{q-1}-1} \left[1-(1-q)\beta H\right]d\nu\cr\cr
&=&\frac{1}{\left[1-(1-q)\beta H\right]^{\frac{1}{1-q}}}\int_{0}^{\infty}e^{-\nu}e^{\nu\left(1-q\right)\beta H} \nu^{\frac{1}{q-1}-1}d\nu{\color{blue},}
\eeq
which gives
\beq\label{part007}
Z_{q}&=&\frac{1}{\Gamma\left(\frac{1}{q-1}\right)}\int_{0}^{\infty}e^{-\nu}e^{-\nu\left(q-1\right)\beta H}\nu^{\frac{1}{q-1}-1}d\nu{\color{blue},}
\eeq
where the function $e^{-\nu\left(q-1\right)\beta H}$ is such that its  trace leads  to the  decomposition into product of two different partition functions
\beq
Tr\left\lbrace e^{-\nu\left(q-1\right)\beta H} \right\rbrace
=Z_{\parallel}\left[\nu\beta\left(q-1\right)\right] Z_{\perp}{\left[\nu\beta\left(q-1\right)\right]}, 
\eeq
with
\beq
Z_{\parallel}\left[\nu\beta\left(q-1\right)\right]&=&L_{z}\left[\frac{M}{2\pi \hbar^{2}\beta \nu\left(q-1\right)}\right]^{\frac{1}{2}}{\color{blue},}
\eeq
and 
\beq
Z_{\perp}\left[\nu\beta\left(q-1\right)\right]&=&\frac{V}{L_{z}}\frac{M\omega_{c}}{2\pi\hbar }\frac{e^{-\nu(q-1)\frac{x}{2}}}{1-e^{-\nu\left(q-1\right)x}} \frac{1}{1-e^{\nu(q-1) \frac{x}{2\Omega}\left(\Omega-\omega_{c}\right)}}{\color{blue},}
\eeq
where we have considered  $x=\beta\hbar\Omega$.
Then (\ref{part007}) writes
\beq
Z_{q}&=&\frac{1}{\Gamma\left(\frac{1}{q-1}\right)}V\left[\frac{M}{2\pi \hbar^{2}\beta \left(q-1\right)}\right]^{\frac{1}{2}}\int_{0}^{\infty}e^{-\nu} \nu^{\frac{1}{q-1}-1}\nu^{-\frac{1}{2}}\cr\cr
&&\times \frac{e^{-\nu(q-1)\frac{x}{2}}}{1-e^{-\nu\left(q-1\right)x}}\frac{1}{1-e^{\nu(q-1) \frac{x}{2\Omega}\left(\Omega-\omega_{c}\right)}}d\nu{\color{blue};}\cr\cr
Z_{q}&=&\frac{1}{\Gamma\left(\frac{1}{q-1}\right)} V \frac{M\omega_{c}}{2\pi \hbar }\left(\frac{M}{2\pi \hbar^{2}\beta }\right)^{\frac{1}{2}}\left(\frac{1}{q-1}\right)^{\frac{1}{2}} I_{q}{\color{blue},}
 \eeq
 with $I_{q}$ delivered by
 \beq
 I_{q}  
 =\Gamma\left(\frac{1}{q-1}-\frac{1}{2}\right)\sum_{n=0}^{\infty}\sum_{m=0}^{\infty}\left[1+\left(q-1\right)xn+\left(q-1\right)\frac{x}{2\Omega}\left(\Omega-\omega_{c}\right)m\right]^{-\left( \frac{1}{q-1}-\frac{1}{2}\right)}.
 \eeq
Thereby,  $Z_{q}$ is obtained as
 \beq
 Z_{q}&=&V\left(\frac{M\omega_{c}}{2\pi\hbar}\right)\left(\frac{M}{2\pi\hbar^{2}\beta}\right)^{\frac{1}{2}}\Gamma\left( \frac{1}{q-1},\frac{1}{2}\right)\frac{1}{1-e^{-\left(\frac{3-q}{2}\right)x}} \frac{1}{1-e^{-\left(\frac{3-q}{2}\right)\frac{x}{2\Omega}\left(\Omega-\omega_{c}\right)}}.
 \eeq
 In the limit $q\rightarrow 1$, we get 
 \beq
 Z_{1}\left(\beta\right)&=&V\left(\frac{M\omega_{c}}{2\pi\hbar}\right)\left(\frac{M}{2\pi\hbar^{2}\beta}\right)^{\frac{1}{2}}  \frac{1}{1-e^{-x}} \frac{1}{1-e^{-\frac{x}{2\Omega}\left(\Omega-\omega_{c}\right)}}.
 \eeq

\end{document}